\documentclass[11pt,tightenlines,article,amssymb,amsmath,amsfonts,superscriptaddress,nofootinbib]{revtex4}

\usepackage{epsfig}
\usepackage{graphics}
\usepackage{graphicx}
\usepackage{amsmath,amssymb,mathrsfs}
\usepackage{amsfonts}
\usepackage[usenames,dvipsnames]{xcolor}
\usepackage{wasysym}
\usepackage{times}
\usepackage{mathptmx}
\usepackage{gensymb}
\usepackage{appendix}
\usepackage{listings}
\usepackage{listings}
\usepackage{url}
\usepackage[normalem]{ulem} 
\usepackage{alltt}
\usepackage[colorlinks]{hyperref}
\usepackage{longtable}
\usepackage{enumitem}
\setlist{nosep}

\newcommand{\AEI}{Max Planck Institute for Gravitational Physics (Albert-Einstein-Institute), Am M\"uhlenberg 1, Potsdam-Golm, 14476, Germany}
\newcommand{\Ligo}{LIGO Laboratory, California Institute of Technology, MS 100-36, Pasadena, California 91125, USA}
\newcommand{\TAPIR}{Theoretical Astrophysics, Walter Burke Institute for Theoretical Physics, California Institute of Technology, Pasadena, California 91125, USA}
\newcommand{\Radboud}{Department of Astrophysics/IMAPP, Radboud University Nijmegen, P.O. Box 9010, 6525 GL Nijmegen, The Netherlands }
\newcommand{\CITA}{Canadian Institute for Theoretical Astrophysics, 60 St.~George Street, University of Toronto, Toronto, ON M5S 3H8, Canada}

\newcommand{\ExNR}{\hat e_x}
\newcommand{\EyNR}{\hat e_y}
\newcommand{\EzNR}{\hat e_z}

\newcommand{\tNR}{\theta}
\newcommand{\pNR}{\phi}

\newcommand{\ErNR}{{\hat r}}
\newcommand{\EtNR}{{\hat\theta}}
\newcommand{\EpNR}{{\hat\phi}}

\newcommand{\hpNR}{h_+^{\rm NR}}
\newcommand{\hcNR}{h_{\times}^{\rm NR}}
\newcommand{\nNR}{\hat{n}}

\newcommand{\lNR}{\hat L}

\newcommand{\tGW}{t_{\rm GW}}

\newcommand{\ExS}{{{\hat x}}}
\newcommand{\EyS}{{{\hat y}}}
\newcommand{\EzS}{{{\hat z}}}

\newcommand{\ExW}{\hat X}
\newcommand{\EyW}{\hat Y}
\newcommand{\EzW}{\hat Z}
\newcommand{\hpW}{h_+^{\rm W}}
\newcommand{\hcW}{h_\times^{\rm W}}

\newcommand{\phiRef}{\Phi} 
\newcommand{\meanAnomaly}{\delta} 
\newcommand{\equalref}{\overset{\rm ref}{=}}

\begin{document}

\title{Numerical Relativity Injection Infrastructure}

\author{Patricia Schmidt}
\email{patricia.schmidt@ligo.org}
\affiliation{\Ligo}
\affiliation{\TAPIR}
\affiliation{\Radboud}

\author{Ian W. Harry}
\email{ian.harry@ligo.org}
\affiliation{\AEI}

\author{Harald P. Pfeiffer}
\email{harald.pfeiffer@ligo.org}
\affiliation{\AEI}
\affiliation{\CITA}

\begin{flushright}
LIGO-T1500606-v7
\end{flushright}

\begin{abstract}
\label{sec:abs}
This document describes the new Numerical Relativity (NR) injection infrastructure in the LIGO Algorithms Library (LAL), 
which henceforth allows for the usage of NR waveforms as a discrete waveform approximant
in LAL. With this new interface, NR waveforms provided in the described format can directly be
used as simulated GW signals (``injections'') for data analyses, which include parameter estimation, searches, hardware
injections etc. As opposed to the previous infrastructure, this new interface natively handles sub-dominant modes and waveforms from numerical simulations of precessing binary black holes, making them directly accessible to LIGO analyses. To correctly handle precessing simulations, the new NR injection infrastructure internally transforms the NR data into the coordinate frame convention used in LAL. 
\end{abstract}

\maketitle 

\section{Introduction}
\label{sec:intro}
LIGO reported the first detections of gravitational waves (GW) from merging binary black holes~\cite{Abbott:2016blz, Abbott:2016nmj, TheLIGOScientific:2016pea}.
Such coalescing compact binaries are prime sources of GWs for ground-based interferometric GW detectors (see e.g.~\cite{Sathyaprakash:2009xs}). Whilst for low-mass systems only the early part of the binary evolution, the inspiral, is accessible to LIGO, for high-mass systems 
also the later stages, in particular the merger and ringdown of the final black, are visible in LIGO's sensitivity band. 

During the early part of the binary coalescence, the emitted gravitational waveforms are accurately described by
analytic post-Newtonian (PN) expansions of the Einstein field equations (see e.g.~\cite{lrr-2014-2}). 
To obtain the waveforms through the final stages of the binary coalescence, the full non-linear solutions
of the field equations are required, which are provided by Numerical Relativity (NR)~\cite{Pretorius:2005gq, Baker:2005vv, Campanelli:2005dd} (see for example~\cite{Centrella:2010mx} for a comprehensive overview).

Numerical Relativity already plays a crucial role in GW data analysis: the construction of waveform models that govern the complete inspiral-merger-ringdown signal (see~\cite{Ohme:2011rm} for a review) depend heavily on NR simulations. Such waveform models~\cite{Hannam:2013oca, Pan:2013rra, Taracchini:2013rva, Khan:2015jqa} underpin LIGO's GW searches, parameter estimation and tests of general relativity~(see \cite{Abbott:2016blz} and references therein). Furthermore, numerical simulations are crucial to determine the remnant black hole's mass and spin~\cite{Healy:2014yta}, and to investigate systematic biases due to waveform modeling errors~\cite{Abbott:2016wiq}.

In the Advanced detector era, it is very advantageous to be able to directly use NR waveforms in gravitational-wave searches and parameter estimation, to test General Relativity and to assess the systematics of analytic waveforms models within a uniform framework. Such use of NR waveforms should be easy and convenient.
This is the purpose of the new \emph{Numerical Relativity Injection Infrastructure} described in this document. Once the NR data is provided in a specific format (as described below), this infrastructure allows for the treatment of
NR waveforms as a ``discrete'' waveform approximant, which can seamlessly be called from within the LIGO Algorithm Library (LAL)\footnote{Available at \url{https://wiki.ligo.org/DASWG/LALSuite}}.

In previous efforts, binary-black-hole (BBH) hybrid waveforms constructed by combining a PN inspiral with an NR merger-ringdown waveform, were used in LIGO data analysis and parameter estimation in the NINJA and NINJA-2 projects~\cite{Aylott:2009ya, Aasi:2014tra}.  
However, the previously employed NR modules in LAL
require the NR waveforms to be resampled at a uniform time-spacing. In the NINJA framework, the resampling was performed \emph{before} inserting the total mass scale. For the waveforms to be
useable at high total mass, the time-spacing in the NR data has to be very small, resulting in very large storage requirements, even if only the dominant harmonics $(\ell,m)=(2,\pm 2)$ were considered.

The new infrastructure described here improves on the earlier approaches
in several significant ways. First, data is stored in a highly efficient compressed format~\cite{Galley:2016mvy}; even including a large number of sub-dominant modes (as is now encouraged), storage requirements are lower than just for the dominant modes in the preceding storage format.  
Second, the compressed NR data are interpolated with one-dimensional spline interpolation \emph{after} the mass scale is inserted. This avoids high-memory operations and further reduces storage requirements and I/O times. Third, the new infrastructure handles projection of the NR data onto arbitrary source-location and detector orientation using the same conventions in LAL as for other waveform families entirely agnostic of the NR code that was used to produce the simulation. The new injection infrastructure is fully implemented in LAL and is intended to supersede previously used NR modules.

The remainder of this technical document is organised as follows: In Sec. \ref{sec:format} we provide a brief summary
of the NR data format and metadata required as input. In Sec. \ref{sec:gen} we describe the basics of the waveform evaluation code and give explicit examples of how the NR waveforms are evaluated in \texttt{lalsimulation}. Sec. \ref{sec:coordinates} details the frame transformations between the NR frame and the LAL wave-frame. We highlight caveats and desired future improvements in Sec.~\ref{sec:discussion}.

\section{Waveform format}
\label{sec:format}
All data for \emph{one} NR simulation is provided in a single HDF5-file.
  This file contains:
  \begin{enumerate}
    \item The gravitational waveforms given as spherical-harmonic
      modes in a spline-compressed format.
    \item Metadata describing the simulation, and identifying the
      origin of the simulation.
      \item Optionally, additional information about the dynamics of
        the black holes.
  \end{enumerate}
When multiple NR datasets for the
identical physical configuration (e.g. during a convergence test, or
for different choices during GW extraction) are provided, then each NR dataset 
including metadata needs to be stored in a separate .h5 file with a unique name.

\subsection{NR conventions}
\label{sec:NRconv}
In Numerical Relativity one solves for the complete space-time of the
binary system. For GW data analysis purposes one requires the
gravitational-wave strain $h$ far from the source. The relevant
numerical quantity is the metric perturbation $h_{ij}$ as computed in
the transverse-traceless (TT) gauge.

There are different ways of computing the metric perturbation from a numerical evolution. The most common methods include the use of the complex Weyl scalar $\Psi_4$~\cite{Newman:1961qr, Penrose:1962ij}, which is related to the metric perturbation via two time derivatives, or the Regge-Wheeler-Zerilli formalism~\cite{Regge:1957td, Zerilli:1970se, Zerilli:1971wd, Moncrief:1974am}, which computes the metric perturbation in the wave-zone as a perturbation of the Schwarzschild spacetime. 

In the TT gauge, the metric perturbation
has two independent real polarisations, $h_+$ and $h_\times$, which can be written as the complex strain
\begin{equation}
\label{wq:Hlm}
h = h_+ - i h_\times \in \mathbb{C},
\end{equation}
where $h_+, h_\times \in \mathbb{R}$. 

Let $( \ExNR, \EyNR, \EzNR)$ be a Cartesian coordinate system in the \emph{wave-zone}, i.e. the zone far away from the binary where the GWs are extracted. This Cartesian coordinate system is related to the polar coordinates $(r, \theta, \phi)$ by the standard transformation. In this coordinate system, henceforth referred to as the NR frame, the metric perturbation is commonly decomposed into \emph{modes}
in a basis of spin-weighted spherical harmonics, ${}^{-2}Y_{\ell m}$, of spin weight $s=-2$, where the GW propagation direction is the radial unit vector $\hat{r}$.
For any point $(\theta, \phi)$ on the unit sphere, the GW strain takes the form
\begin{equation}
\label{ }
h^\mathrm{NR}(\tGW; \theta, \phi) = h^\mathrm{NR}_+ - i h^\mathrm{NR}_\times = \sum_{\ell=2}^\infty \sum_{m=-\ell}^{\ell} H_{\ell m}(\tGW) {}^{-2}Y_{\ell m}(\theta,\phi),
\end{equation}
where $H_{\ell m}(\tGW)$ denotes the extracted NR gravitational-wave modes.
The modes are given in terms of a retarded time-coordinate
  $t_{\rm GW}$.
As for any wave, we can also write each mode $H_{\ell m}(\tGW)$ as an amplitude $A_{\ell m}(\tGW)$ and a phase
$\Phi_{\ell m}(\tGW)$,
\begin{equation}
\label{ }
H_{\ell m}(\tGW) = A_{\ell m}(\tGW)e^{i\Phi_{\ell m}(\tGW)}.
\end{equation}
The mode amplitude $A_{\ell m}(t)$ is defined as the complex norm of the complex time series $H_{\ell m}(\tGW)$, the phase $\Phi_{\ell m}(\tGW)$ is the unwrapped argument of the complex time series $H_{\ell m}(\tGW)$. For a binary that is orbiting counter-clockwise in the xy-plane of the NR frame (i.e., the orbital angular frequency vector is \emph{parallel} to the z-axis), $\Phi_{2,2}(\tGW)$ is a monotonically {\bf decreasing} function.

Following the LAL waveform convention, the time-coordinate $\tGW$ in the waveform modes has to be chosen such that the peak of the waveform occurs at $t_\mathrm{GW,peak} \equiv 0$, where the ``peak'' of the waveform is defined by
\begin{equation}
\label{eq:peak}
h_\mathrm{peak} := \max \left( \sum_{\ell, m} A_{\ell m}(\tGW)^2 \right).
\end{equation}

Note that the time coordinate $\tGW$ in the wave-zone is not
  identical to time coordinate $t_\mathrm{NR}$ used by the numerical
  relativity simulation in the strong-field regime, the latter is used
  below to give information about the dynamics of the black holes.  
  There is no unambiguous identification
  of $t_{\rm GW}$ with $t$, since they are defined in
  different regions of the space-time (far-zone vs. near-zone).  The
  normalization of $t_{\rm GW}$ is given by Eq.~(\ref{eq:peak}); we
  describe on page~\pageref{tGW-vs-tNR} how to choose $t$.

\subsection{Amplitude and phase spline compression}
\label{sec:spline}
Gravitational waveforms for LIGO data analysis purposes require
uniform sampling in time for a given sampling frequency.  NR datasets,
however, are commonly not uniformly sampled and if they are, the
sampling interval $\Delta \tGW$ may not necessarily correspond to the
one required by data analysis tools.  It is therefore unavoidable to
interpolate the NR data to the desired sampling rate. Whilst the NR
data could simply be interpolated as they are, we choose to reduce the
data by performing one-dimensional spline compression on the NR
data~\cite{Galley:2016mvy}. This is a particular advantage for long
simulations or hybrid data, but also significantly reduces the storage
and I/O for pure NR data.

The one-dimensional spline compression is performed separately for each {\bf mode amplitude} $A_{\ell m}(\tGW)$ and {\bf mode phase} $\Phi_{\ell m}(\tGW)$, which are
already time-shifted such that $\tGW=0$ corresponds to the peak of the waveform. 
Pure NR data without an inspiral need to have the initial \emph{junk radiation} removed \emph{before} the spline interpolants are constructed. We refer to this very first data point stored in the time series after removing the initial junk as the \emph{beginning} of the waveform. No tapering should be applied at the beginning or end of the waveform data. 

The routine employed to compress and interpolate the data is the \emph{reduced-order spline interpolation} presented in~\cite{Galley:2016mvy}. It uses a greedy algorithm that selects the near-optimal points to construct a univariate spline interpolant with a specified global accuracy and polynomial degree.
By default, the interpolants are constructed using fifth degree polynomials and a tolerance of $10^{-6}$,
i.e., if the spline is evaluated at the original discrete NR times $\tGW$, the original NR values for mode amplitude and phase are recovered with an error equal to or smaller than the specified tolerance. In addition, a median error can be associated with any predicted value, i.e., a value not contained in the original NR data. 
For a detailed description of this method and the accuracy of the obtained interpolants, we refer the reader to~\cite{Galley:2016mvy}. The spline compression is conveniently performed using the publicly available Python package \texttt{romSpline} by Chad R. Galley~\cite{romspline}. The spline interpolants for each amplitude and phase are obtained via \verb"romSpline" as follows: 
\begin{alltt}
import romSpline 
spline = romSpline.ReducedOrderSpline(\(\tGW\), \(A\sb{\ell,m}(\tGW) or \Phi\sb{\ell,m}(\tGW)\), 
																		                                     verbose=False)
spline.write('filename.h5')
\end{alltt}
The output of \texttt{spline.write} contains all information needed for subsequent interpolation of the respective input time-series data and is composed of five datasets: (deg, tol, X, Y, errors)\footnote{The latest version of {\tt romSpline} allows to pass a group descriptor, so that the spline-data can be written directly into the appropriate group of the final output file.}.

The spline interpolants for each $(\ell,m)$-mode for a single NR
simulation are stored as individual H5-groups in the
\emph{final HDF5 output file} under the mandatory group-names given below and summarized in Sec.~\ref{sec:format:groups}:\\

 \begin{longtable}{|p{4cm}|p{2.5cm}|p{9.6cm}|}
   \hline  \emph{Group-name} & \emph{Type} & \emph{Description} \\ \hline
   \endhead
   \hline  \emph{Group-name} & \emph{Type} & \emph{Description} \\ \hline
 \endfirsthead
 \hline 
  \multicolumn{3}{r}{\emph{continued}}
 \endfoot
 \hline
 \endlastfoot
   \hline 
 \textbf{amp\textunderscore l\#1\textunderscore m\#2} & romSpline-group & amplitude $A_{\ell m}(t_{\rm GW})$ \\
 \textbf{phase\textunderscore  l\#1\textunderscore m\#2} & romSpline-group  & phase $\Phi_{\ell m}(t_{\rm GW})$
 \\ \hline 
 \end{longtable}
 $ $ \\
 Here (\#1, \#2) are placeholders for $(\ell, m)$, e.g., for $(\ell=2, m=-2)$ the group naming convention is phase\textunderscore l2\textunderscore m-2 and amp\textunderscore l2\textunderscore m-2. 
 
\subsection{Metadata stored as HDF5-attributes}
\label{sec:meta}
The metadata format is adapted from the original NINJA-2 metadata format~\cite{Brown:2007jx}. The metadata are described in the following list, and are stored as \emph{attributes} of the final HDF5 file for each NR simulation. Often, more extensive metadata are available for a NR simulation. We recommend that such additional metadata is included as a free-format text dataset in the 'auxiliary-info' H5-group (see below). 

Required time-dependent metadata (see list below) have to contain the values corresponding to the first entry in the stored time series. For pure NR data, since the junk radiation has to be removed, these are \emph{not} the initial data of the simulation. We refer to this very first data point stored in the time series as the \emph{beginning} of the waveform data, subsequently indicated by $t_\mathrm{begin}$. 

Vectors must be represented by their Cartesian components in the NR frame (see Sec.~\ref{sec:NRconv}). This coordinate system is consistent with the one used in the decomposition in spherical harmonics (e.g. the polar axis of the spherical harmonics must point along the +z-axis). No requirements are placed on the orientation of the NR frame. For instance, the orbital unit separation vector 'nhat' and the Newtonian unit orbital angular momentum 'LNhat' need not point along specific coordinate axes. Note also that the spin-fields in the metadata 'spin\{1,2\}\{x,y,z\}' are given in the NR frame, and \emph{not} in the LAL frame. The rotation of the GW modes into the LAL frame is taken care of by the NR injection infrastructure, following the conventions described in Sec.~\ref{sec:coordinates}. 

Note: The time-coordinate of spins, positions, masses and their
respective time-series will be associated with the apparent horizon,
e.g. the NR coordinate time $t$. Such a time cannot
unambiguously be identified with the retarded time $\tGW$ of the
waveform modes.
\label{tGW-vs-tNR}
However, the submitter is asked to make a {\bf reasonable
effort to have the same numerical values of $t$ and $\tGW$
to correspond to the same portion of the waveform}, e.g. through
choosing $\tGW = t - R^*_\mathrm{extraction}$, where
$R^*_\mathrm{extraction}$ is the tortoise radius of the extraction sphere~\cite{Fiske:2005fx,Boyle:2009vi}. \\
\newline

\noindent

\renewcommand*{\arraystretch}{1.5} 
\begin{longtable}{|p{3.4cm}|p{1.6cm}|p{11.2cm}|}
  \hline  \emph{Attribute-name} & \emph{Type} & \emph{Description} \\ \hline
  \endhead
  \hline  \emph{Attribute-name} & \emph{Type} & \emph{Description} \\ \hline
\endfirsthead
\hline 
 \multicolumn{3}{r}{\emph{continued}}
\endfoot
\hline
\endlastfoot
  \textbf{Format} & integer & indicates what data are supplied. Must be 1, 2, 3\\
  \textbf{type} & string & keyword description of the simulation as requested by the LIGO Open Science Center (LOSC). Required value: NRinjection\\
  \textbf{name} & string  & short identifier of the simulation, e.g., SXS:BBH:0019\\

\textbf{alternative-names} & string & comma-separated list of user-specifiable alternative names. These names can be longer, more descriptive, and/or include what specific series of simulations this configuration belongs to.\\

\textbf{NR-group} & string &  name of the NR group that carried out the simulation\\

\textbf{NR-code} & string &  name of the NR code that was used to carry out the simulation\\

\textbf{modification-date} & string & date when this .h5 file was last updated. Format 'YYYY-MM-DD'\\

\textbf{point-of-contact-email} & string & contact person for questions\\

\textbf{simulation-type} & string & keyword description of the spin configuration. Allowed values are: aligned-spins, non-spinning, precessing\\

\textbf{INSPIRE-bibtex-keys} & string & comma-separated list of INSPIRE bibtex keys that should be cited when this waveform is used (1-3 publications)\\

\textbf{license} & string & allowed values are: LVC-internal, public\\

\textbf{Lmax} & integer & the maximum $\ell$-value for which  $h_{\ell m}$-modes are supplied (all modes with $\ell \leq \mathrm{Lmax}$, $-\ell \leq m \leq +\ell$ have to be provided)\\

\textbf{NR-techniques} & string & attempts to summarize major elements of the NR simulation. A comma-separated list of one element in each of the following categories \\
  & & \parbox{11.2cm}{
Category 1: Puncture-ID, Quasi-Equilibrium-ID \\
Category 2: BSSN, GH, Z4c \\
Category 3: RWZ-h,  Psi4-integrated \\
Category 4: Finite-Radius-Waveform, CCE-Waveform, Extrapolated-Waveform \\
Category 5: ApproxKillingVector-Spin, CoordinateRotation-Spin \\
Category 6: Christodoulou-Mass \\
Example: 'Quasi-Equilibrium-ID, GH, RWZ-h, Extrapolated-Waveform, ApproxKillingVector-Spin, Christodoulou-Mass'\\
\emph{(Note: This list is extensible. If your code does not fit the given choices, contact the authors.)}
}\\
\textbf{files-in-error-series} & string &  a comma-separated list of .h5 files (including the present one) that combined form an error series for the binary configuration, e.g. different numerical resolutions. Set to ' ' if no error-series for this configuration exists.\\

\textbf{\footnotesize comparable-simulation} & string & one other .h5 file that (a) has an error-series and (b) is numerically ``comparable'' to the present one, i.e. an error-analysis that is performed on 'comparable-simulation' is expected to carry over to this waveform. Set to ' ' if an error-series is provided.\\

\textbf{production-run} & integer & allowed values are 1 and 0. If 1, this is the highest quality member of the error-series and should be used for analyses.  If 0, this is a lower-quality member of the error-series and should not be used for general analyses.\\

\textbf{object1} & string & keyword description to identify the object type. Allowed values are: BH, NS\\

\textbf{object2} & string & keyword description to identify the object type. Allowed values are: BH, NS\\

\textbf{mass1} & float & mass of the more massive object at $t_{\rm begin}$; if both objects are BH, the unit of mass is arbitrary. If at least one object is a NS, then the unit is solar mass $M_\odot$.\\

\textbf{mass2} & float & mass of the lighter object at $t_{\rm begin}$; if both objects are BH, the unit of mass is arbitrary. If at least one object is a NS, then the unit is solar mass $M_\odot$.\\

\textbf{eta} & float & the symmetric mass ratio of the simulation at $t_{\rm begin}$.\\

\textbf{f\textunderscore lower\textunderscore at\textunderscore 1MSUN} & float & frequency of the $(2,2)$-mode in Hz at the beginning of the waveform scaled to $1M_\odot$\\

\textbf{spin1x} & float & x-component of the dimensionless spin vector $\vec\chi_1(t_{\rm begin})$ in NR frame\\

\textbf{spin1y} & float & y-component of the dimensionless spin vector $\vec\chi_1(t_{\rm begin})$ in NR frame\\

\textbf{spin1z} & float & z-component of the dimensionless spin vector $\vec\chi_1(t_{\rm begin})$ in NR frame\\

\textbf{spin2x} & float & x-component of the dimensionless spin vector $\vec\chi_2(t_{\rm begin})$ in NR frame\\

\textbf{spin2y} & float & y-component of the dimensionless spin vector $\vec\chi_2(t_{\rm begin})$ in NR frame\\

\textbf{spin2z} & float & z-component of the dimensionless spin vector $\vec\chi_2(t_{\rm begin})$ in NR frame\\

\textbf{LNhatx} & float & x-component of the Newtonian orbital angular momentum unit vector $\hat L_N(t_{\rm begin})$ in NR frame\\

\textbf{LNhaty} & float & y-component of the Newtonian orbital angular momentum unit vector $\hat L_N(t_{\rm begin})$ in NR frame\\

\textbf{LNhatz} & float & z-component of the Newtonian orbital angular momentum unit vector $\hat L_N(t_{\rm begin})$ in NR frame\\

\textbf{nhatx} & float & x-components of the orbital separation unit vector $\hat n(t_{\rm begin})$ given by Eq.~(\ref{eq:nhat}) in NR frame\\

\textbf{nhaty} & float & y-components of the orbital separation unit vector $\hat n(t_{\rm begin})$ given by Eq.~(\ref{eq:nhat}) in NR frame\\

\textbf{nhatz} & float & z-components of the orbital separation unit vector $\hat n(t_{\rm begin})$ given by Eq.~(\ref{eq:nhat}) in NR frame\\

\textbf{Omega} & float & dimensionless orbital frequency $M\Omega(t_{\rm begin})$ \\

\textbf{eccentricity} & float & estimated eccentricity of the simulation at $t_{\rm begin}$\\

\textbf{mean\textunderscore anomaly} & float & estimated mean anomaly
(cf. Eq.~\ref{eq:anomaly}) at $t_{\rm begin}$. For
  $\mathrm{eccentricity} \leq 10^{-3}$, it is allowed to set
  mean\textunderscore anomaly to 0. If the mean anomaly has not been
  computed, set mean\textunderscore anomaly to
  -1.\\ \textbf{PN\textunderscore approximant} & string & \emph{Only
    present for PN-NR hyrbid waveforms}: identifier of the inspiral
  approximant
\end{longtable}

\subsection{Data stored as HDF5-groups and datasets}
\label{sec:format:groups}
The following groups are required inside the .h5 file that represents
a simulation.
Some groups are optional and only need to be given if Format=2 or Format=3.
All time-series represent ROM-compressed data obtained via
\texttt{romSpline} in the same way as the amplitudes and phases (see
Sec.~\ref{sec:spline} for details). \\

\begin{longtable}{|p{4cm}|p{2.5cm}|p{9.6cm}|}
  \hline  \emph{Group-name} & \emph{Type} & \emph{Description} \\ \hline
  \endhead
  \hline  \emph{Group-name} & \emph{Type} & \emph{Description} \\ \hline
\endfirsthead
\hline 
 \multicolumn{3}{r}{\emph{continued}}
\endfoot
\hline
\endlastfoot
\textbf{auxiliary-info} & H5-group & Contains anything that the submitter finds helpful to identify, document and repeat the run\\ 
\textbf{NRtimes} & H5-dataset & \emph{optional but highly recommended:} 1-d array of the discrete times $t_{\rm GW}$ that formed the input-times into the romSpline compression of
 {\tt amp\textunderscore l\#1\textunderscore m\#2} and {\tt phase\textunderscore \#1\textunderscore m\#2} \\ \hline

\multicolumn{3}{|l|}{
  \rule[-.7em]{0pt}{2em}
       {\bf GW modes:} \emph{Two groups for each \boldmath$(\ell, m)$, $2\le \ell\le L_{\rm max}$, $-\ell\le m\le +\ell$}}\\ \hline
\textbf{amp\textunderscore l\#1\textunderscore m\#2} & romSpline-group & amplitude $A_{\ell m}(t_{\rm GW})$  \\
\textbf{phase\textunderscore  l\#1\textunderscore m\#2} & romSpline-group  & phase $\Phi_{\ell m}(t_{\rm GW})$
\\ \hline

\multicolumn{3}{|l|}{
  \rule[-.7em]{0pt}{2em}
       {\bf iI Format \boldmath $\mathbf{\ge 2}$, also specify the following time-series:}}\\ \hline
\textbf{mass1-vs-time} & romSpline-group & mass of the more massive object \\
\textbf{mass2-vs-time} & romSpline-group & mass of the less massive object \\
\textbf{spin1x-vs-time} & romSpline-group & x-component of the dimensionless spin $\vec\chi_1(t)$\\
\textbf{spin1y-vs-time} & romSpline-group & y-component of the dimensionless spin $\vec\chi_1(t)$\\
\textbf{spin1z-vs-time} & romSpline-group & z-component of the dimensionless spin $\vec\chi_1(t)$\\
\textbf{spin2x-vs-time} & romSpline-group & x-components of the dimensionless spin $\vec\chi_2(t)$ \\
\textbf{spin2y-vs-time} & romSpline-group & y-components of the dimensionless spin $\vec\chi_2(t)$\\
\textbf{spin2z-vs-time} & romSpline-group & z-components of the dimensionless spin $\vec\chi_2(t)$\\
\textbf{position1x-vs-time} & romSpline-group & x-component of the center $\vec c_1(t)$ of object1 \\
\textbf{position1y-vs-time} & romSpline-group & y-component of the center $\vec c_1(t)$ of object1\\
\textbf{position1z-vs-time} & romSpline-group & z-components of the center $\vec c_1(t)$ of object1\\
\textbf{position2x-vs-time} & romSpline-group & x-component of the center $\vec c_2(t)$ of object2\\
\textbf{position2y-vs-time} & romSpline-group & y-component of the center $\vec c_2(t)$ of object2\\
\textbf{position2z-vs-time} & romSpline-group & z-component of the center $\vec c_2(t)$ of object2\\
\textbf{LNhatx-vs-time} & romSpline-group & x-component of Newtonian angular momentum direction $\hat L_N(t)$\\
\textbf{LNhaty-vs-time} & romSpline-group & x-component of Newtonian angular momentum direction $\hat L_N(t)$\\
\textbf{LNhatz-vs-time} & romSpline-group & x-component of Newtonian angular momentum direction $\hat L_N(t)$\\
\textbf{Omega-vs-time} & romSpline-group & dimensionless orbital frequency $M\Omega(t)$ \\ 

\hline

\multicolumn{3}{|l|}{
  \rule[-.7em]{0pt}{2em}
       {\bf If Format \boldmath $\mathbf{\ge 3}$, also specify the following time-series}}\\ \hline

\textbf{remnant-mass-vs-time} & romSpline-group &  remnant mass\\
\textbf{remnant-spinx-vs-time} & romSpline-group & x-component of the dimensionless spin of the remnant \\
\textbf{remnant-spiny-vs-time} & romSpline-group & y-component of the dimensionless spin of the remnant \\
\textbf{remnant-spinz-vs-time} & romSpline-group & z-component of the dimensionless spin of the remnant \\
\textbf{remnant-positionx-vs-time} & romSpline-group & x-component of the center of the remnant \\
\textbf{remnant-positiony-vs-time} & romSpline-group & y-component of the center of the remnant \\
\textbf{remnant-positionz-vs-time} & romSpline-group & z-component of the center of the remnant 
\end{longtable}

\section{NR waveform evaluation in LAL}
\label{sec:gen}
Once the HDF5 file has been provided, the NR waveforms can be evaluated through the standard waveform interfaces \texttt{ChooseTDWaveform} in LAL. The approximant name is ``NR\textunderscore hdf5''. 
The spline data are read from file and evaluated for the desired extrinsic parameters, total mass and starting frequency. Since some intrinsic parameters of an NR simulations are fixed (e.g. mass-ratio and dimensionless spins), internal checks on the mass ratio and the spin components are performed to guarantee the consistency between the values passed in the waveform generation call and metadata values. 
Note: The waveform generator requires the input spin values to be defined as given by Eqn.~(\ref{eq:SpinConsistency}). In general, these are different to the values of the spin metadata and need to be computed using the metadata for the spins, the orbital angular momentum and the orbital separation. The function \texttt{SimInspiralNRWaveformGetSpinsFromHDF5File} returns the spins the required convention. 

For a given starting frequency and total mass, a time array is allocated based on an estimate of the waveform length. We use the LAL-function 
\texttt{SimIMRSEOBNRv2ChirpTimeSingleSpin} to estimate the waveform length with an additional leverage of 10\%. If the NR waveforms are not long enough for a given total mass and starting frequency, the generation is aborted and an error is generated. From the estimated length and the desired sampling rate, the discrete time series for the spline evaluation is determined.

To construct the NR GW polarisations $h_+$ and $h_\times$ in the LAL
wave-frame, first the splines for each NR amplitude and phase are
first evaluated at the required sampling times and convolved with the
spin-weighted spherical harmonics
Finally, the NR polarizations are transformed into the LAL wave-frame following Eqs.~(\ref{eq:PolarizationTrafo}). Note that all $(\ell, m)$-modes present in the HDF5 file are used
to compute the two polarisations. 

The compressed NR data files do not store the splines themselves, but the X-data, Y-data, errors, the polynomial degree etc.  A regular GSL interpolation
is used to construct the splines from the HDF5 file. A comparison with the
scipy function \texttt{UnivariateSpline} found that the mismatch between
waveforms reconstructed using the two different interpolators was less than
$10^{-7}$. This is consistent with the level of disagreement expected due to the different numerical interpolation routines. Fig.~\ref{fig:waveforms}
shows an example comparison between NR waveforms obtained using the two
different interpolation routines.
The source code can be found in \texttt{lalsuite/lalsimulation/src/LALSimIMRNRWaveforms.c}.

\subsection{Examples}
\begin{figure}
\begin{center}
\includegraphics[width=80mm]{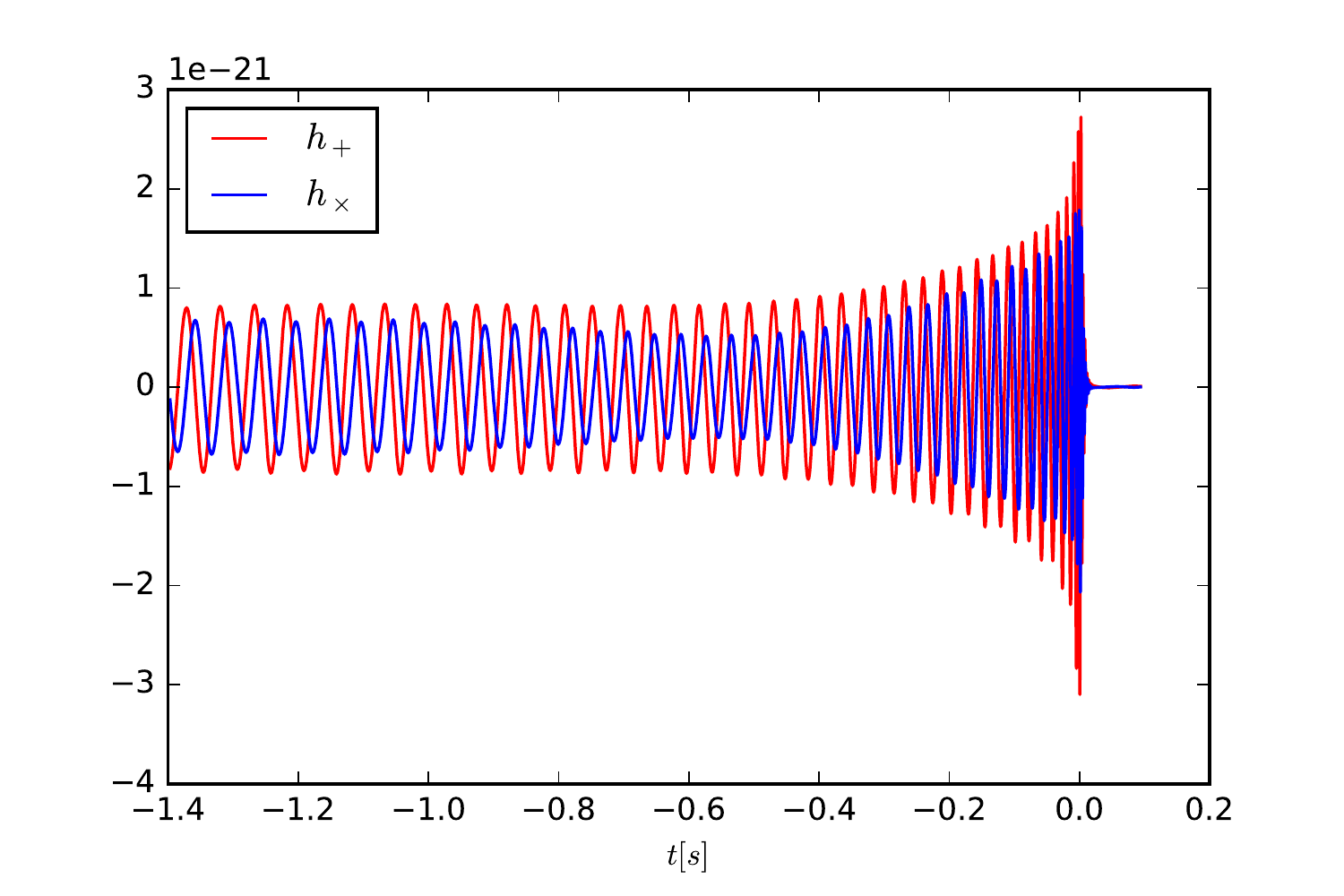}
\includegraphics[width=80mm]{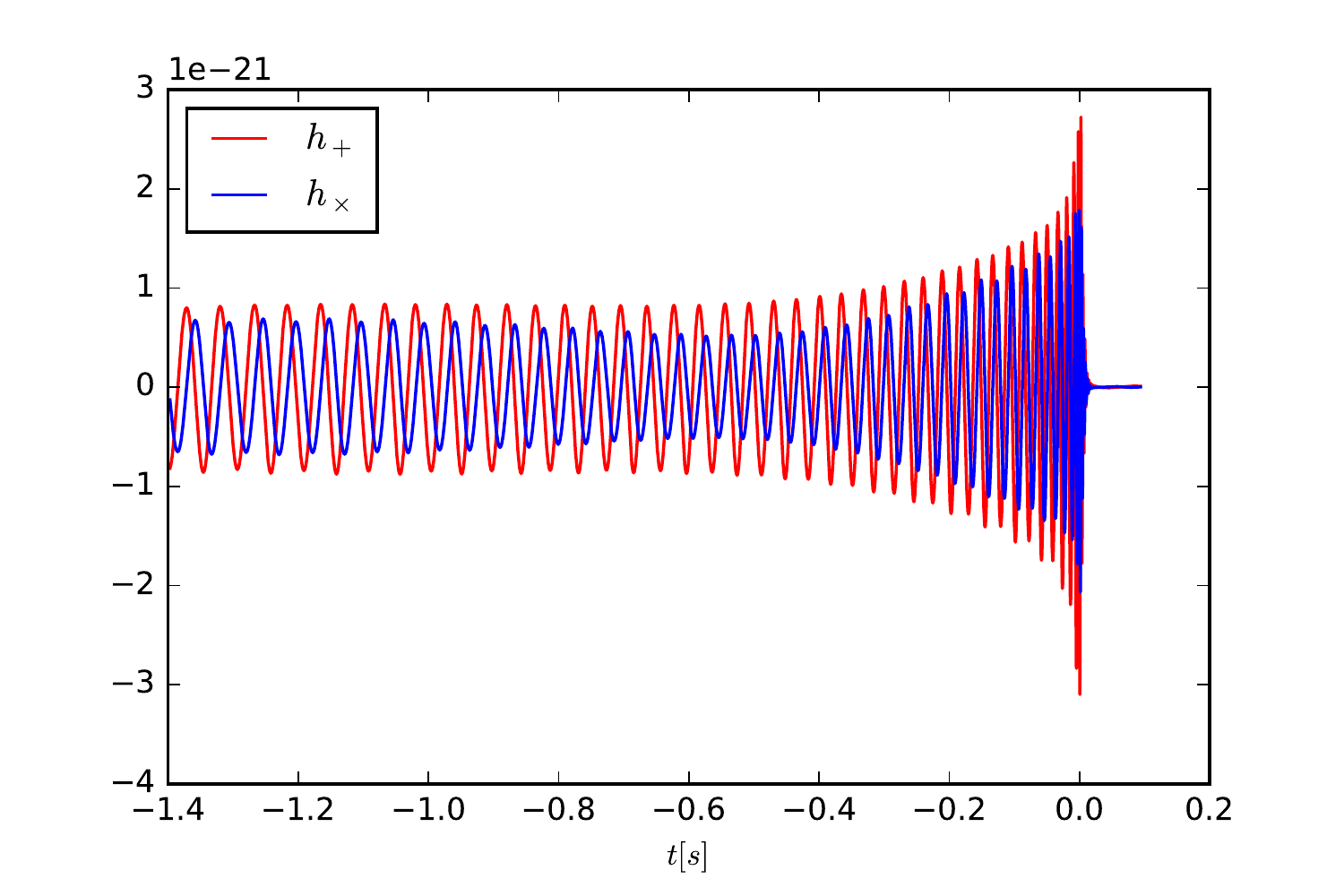}
\caption{The waveform polarizations $h_+$(red) and $h_\times$(blue) of the publicly available SXS waveform SXS:BBH:0006
generated via the waveform interface in \texttt{lalsimulation} (left panel)
and constructed using scipy's \texttt{UnivariateSpline} function(right panel). }
\label{fig:waveforms}
\end{center}
\end{figure}

There are a variety of different ways to evaluate NR waveforms using LIGO data analysis software. Here, we give an explicit example
using Python and the SWIG-wrapped version of \texttt{lalsimulation}.
The only difference between this and generating a waveform using any other
waveform model is that the path to the HDF5 file must be provided explicitly,
as illustrated. \\
Example using \texttt{lalsimulation} through SWIG:
\begin{alltt}
import lal
import lalsimulation as lalsim
# Compute spins in the LAL frame
s1x, s1y, s1z, s2x, s2y, s2z = \newline lalsim.SimInspiralNRWaveformGetSpinsFromHDF5File('/PATH/TO/H5File')
# Create a dictionary and pass /PATH/TO/H5File
params = lal.CreateDict()
lalsim.SimInspiralWaveformParamsInsertNumRelData(params, '/PATH/TO/H5File')
# Generate GW polarisations 
hp, hc = lalsim.SimInspiralChooseTDWaveform(mass1 * MSUN_SI, 
	              mass2 * MSUN_SI,
              s1x, s1y, s1z,
              s2x, s2y, s2z, 
              distance, inclination,
              phiRef, $pi/2$,
              0., 0., deltaT, 
              fStart, fRef,
              params, approximant=lalsim.NR_hdf5)
\end{alltt}
Fig.~\ref{fig:waveforms} shows the two waveform polarizations $h_+$ and $h_{\times}$ for the precessing binary black hole hole simulated in case 
SXS:BBH:0006 from the publicly available SXS catalogue~\cite{Mroue:2013xna} for a total mass of 50 $\mathrm{M}_\odot$ and an inclination of $\pi/3$. The dimensionless spins for this simulation in the LAL frame are 
$\vec{\chi}_1=(-0.05, -0.27, -0.16)$ and $\vec{\chi}_2=(-0.02, -0.11, -0.10)$ and the component masses are $m_1=28.68$ and $m_2=21.32$. The waveform is generated from its beginning, corresponding to the starting frequency of fStart=18.76Hz.
Further parameters are: distance=100Mpc, deltaT=1.0/16384, phiRef=0 and fRef=fStart.

\section{LAL coordinate frames for precessing binaries and NR injections}
\label{sec:coordinates}

\subsubsection{Executive summary: changes of conventions}

It has recently became apparent that certain waveform conventions in LAL
are not ideal to specify precessing binaries: (i) The
phase-angle {\tt phiRef} couples the specification of the line-of-sight to
Earth with the specification of spin components {\tt S1x, S1y, S1z, S2x,
S2y, S2z}. To compute waveforms for the \emph{identical} compact binary
viewed from different directions, one may have to specify different
values for the spin-components {\tt S1x, S1y, S2x, S2y}.
(ii) Several semi-analytic waveform models do not conform to this 
convention, already following the convention detailed below.

Concretely, the spin-components are now specified in a geometric
way based on the angular momentum $\lNR$ and the line connecting object 2 to
object 1, $\nNR$:
\begin{align}
  \mbox{\tt S1x}&\equalref \vec\chi_1\cdot\nNR,\\
  \mbox{\tt S1y}&\equalref \vec\chi_1\cdot (\lNR\times\nNR),\\
  \mbox{\tt S1z}&\equalref \vec\chi_1\cdot\lNR,
\end{align}
(and similarly for the second object). The symbol $\equalref$ indicates
that equality only holds at a reference time as spins generically
precess during an inspiral~\cite{Apostolatos:1994mx, Kidder:1995zr}.

Furthermore, it is suggested to specify
orientation of eccentric orbits through the angles
\begin{align}
  \mbox{\tt phiRef} = \phiRef &\equalref \angle(\nNR, \mbox{line-of-ascending-node}),\\
  \mbox{\tt mean\_anomaly}= \delta.& 
  \end{align}

\subsubsection{Motivation \& benefits of new conventions}
Generally waveform modeling requires at least two coordinate
systems, a ``source-frame'' in which it is convenient to specify
properties of the source of gravitational waves, and a ``wave-frame''
which is adopted to wave-propagation to GW detectors on Earth.
Furthermore, NR data are specified in whatever
coordinates are employed during the numerical evolution, generally
resulting in a third coordinate system. This section defines coordinate frames 
for use in LAL. Specifically, we
achieve:
\begin{enumerate}
  \item Identification of a set of intrinsic parameters that fully
    describe the dynamics of a binary on an eccentric orbit,
    defined solely in terms of the source-frame (i.e. independent of
    the wave-frame).
  \item Identification of three angles that describe the
    transformation between source- and wave-frame, which are
    independent of the intrinsic parameters.
  \item Identification of parameters describing the orbital phase and
    periapsis location, which have a convenient circular-orbit limit:
    As the eccentricity tends to zero, one of the two phase-parameters
    reduces to the standard orbital phase for circular orbits, whereas
    the other becomes irrelevant.
  \item Identities that relate the basis-vectors in the source-frame to the
    wave-frame (and vice versa).  These identities are written in
    vectorial form and are valid in any coordinate system.
\end{enumerate}

The orthogonal decomposition into intrinsic and extrinsic parameters
(points 1 and 2) allows to change the direction at which a binary is
viewed (i.e. the wave-frame), without having to adjust the parameters
that determine the intrinsic dynamics.  Point 3 prepares the ground
for easy extension to eccentric waveforms.
point 4 is of particular relevance when translating NR data into
LAL-conventions: Evaluating the vector identities in the NR-coordinate
system, yields immediately the relation between NR coordinates and
wave-frame.

\subsubsection{NR waveform injections}

NR data are assumed to be supplied in the data-format defined in
Sec.~\ref{sec:format} and~\ref{sec:gen}, and so it needs to be
transformed into the LAL wave-frame.  For a given reference time, an
entire NR data-set could in principle be transformed into the LAL
frame by suitably transforming each $H_{\ell m}$-mode.  Applying
such a coordinate transformation on the NR-data as a pre-processing step
before using it for injections suffers from two disadvantages:
First, the person who prepares the NR-data for LAL-use must perform
the rotation, and must do so correctly. Since NR data is prepared
separately by several NR groups, this opens the possibility of
introducing errors in this step.  Secondly, for precessing systems the
orbital angular momentum $\lNR$ precesses.  Therefore, pre-rotating
the NR-data locks in the reference point, and one would need to
generate different pre-transformed NR-data for different reference points.
Since NR data is in geometric units, the NR-data would have to be
separately transformed whenever the total mass of an injection changes.

To avoid both disadvantages, it is proposed to leave the NR-data in
its original frame (see Sec.~\ref{sec:format}).  Instead, the transformations between the
LAL-convention and the NR-data are applied \emph{at use} during the
call to the waveform evaluation function in LAL (cf. Sec.~\ref{sec:gen}).

This section develops the necessary transformations to implement this
technique. No assumptions are made on the NR-coordinate system in
order to allow for precession when NR-orbital angular momentum is
typically \emph{not} along the z-axis of the NR-coordinate system.
The derived formulae also allow to choose an arbitrary reference point
for the NR waveform, as long as orbital angular momentum and vector
connecting the two compact objects are known at that point.

\subsection{Coordinate frames}
\label{sec:Frames}

The frames defined here differ somewhat from the preceding LAL
conventions as detailed in~\cite{inspiral}.
The differences are needed to achieve the separation between intrinsic
and extrinsic parameters.  In the old conventions, the orbital phase
$\phi$ was also used as part of the rotation parameters that define
the rotation between source- and wave-frame.  Therefore, to ``look''
at the same binary from different angles used to require a suitable
change in the spin-components tangential to the orbital plane.

\begin{figure}
  \includegraphics[width=80mm]{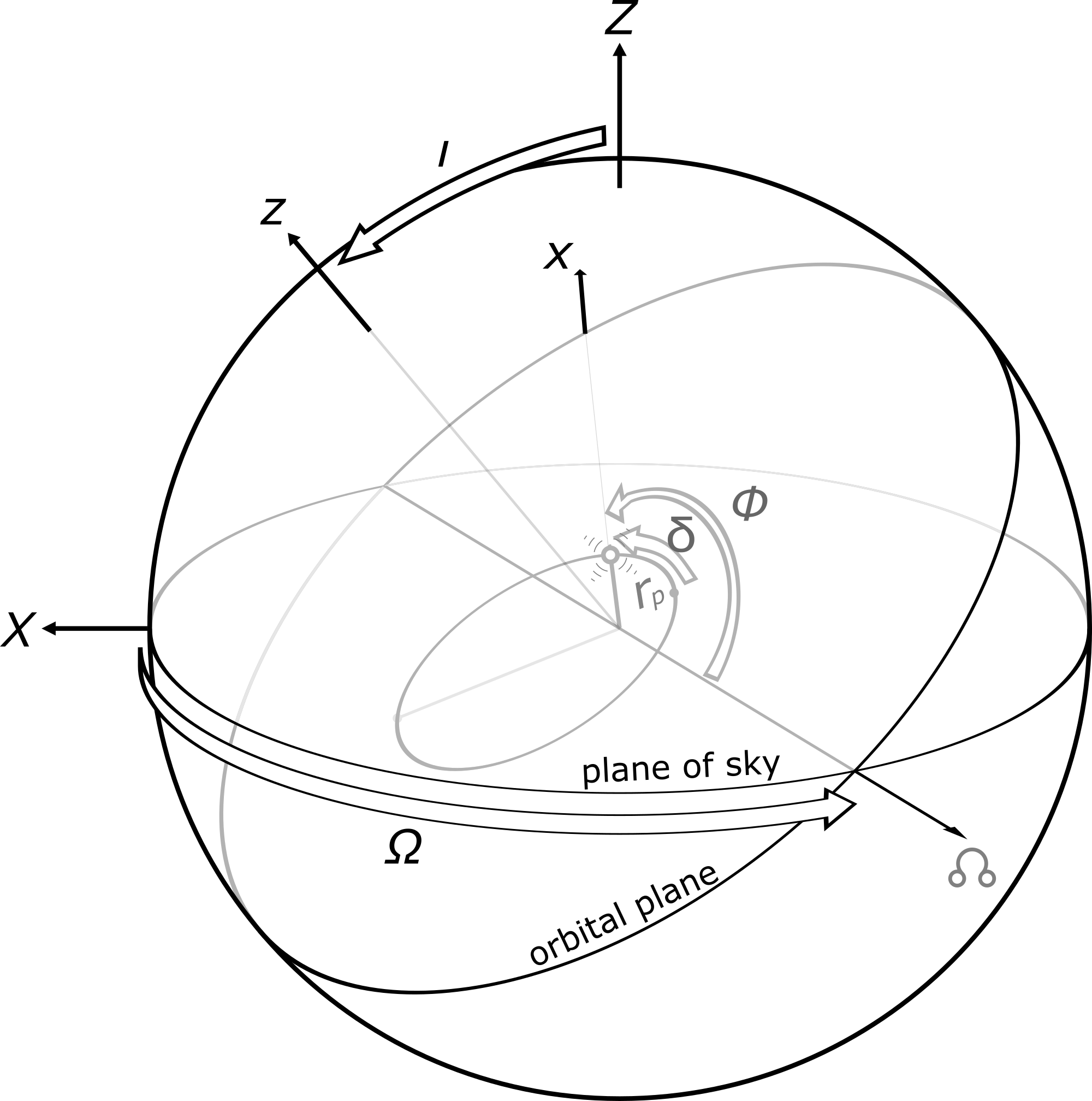}
  \caption{
  \label{fig:frames} Coordinate frames.
  (x,y,z) denotes the source-frame and (X,Y,Z) the wave-frame. For the recommended default
    $\Omega=\pi/2$, the line of ascending nodes agrees with $\EyW$
    i.e. the rotation by the inclination angle $\iota$ is about the
    $\EyW$-axis.}
  \end{figure}

\subsubsection{NR frame \boldmath$(  \ExNR, \EyNR, \EzNR)$}

This is a generic coordinate system without any regards to the
concrete binary motion. Generic coordinate systems occur in numerical
relativity, where the coordinates are chosen through some
gauge-conditions, and the binary is evolving from some initial data.
At some later time therefore, the coordinates will not have any
particular, controlled properties.
The Cartesian basis-vectors are denoted
\begin{equation}
  \ExNR, \EyNR, \EzNR.
\end{equation}
From these, spherical basis-vectors can be computed as
\begin{subequations}
  \label{eq:NRspherical}
\begin{align}
  \ErNR & = \cos\pNR\sin\tNR\;\ExNR + \sin\pNR\sin\tNR\;\EyNR +\cos\tNR\;\EzNR\\
  \EtNR & = \cos\pNR\cos\tNR\;\ExNR + \sin\pNR\cos\tNR\;\EyNR -\sin\tNR\;\EzNR\\
  \EpNR & =        -\sin\pNR\;\ExNR +         \cos\pNR\;\EyNR.
\end{align}
\end{subequations}

NR data are assumed to be represented by spherical-harmonic modes of the
NR coordinates, as previously defined in the NINJA data-formats
document~\cite{Brown:2007jx}:
\begin{subequations}\label{eq:hNR}
\begin{align}
  \hpNR(\tGW) & = \frac{1}{2}\left(\EtNR_i\EtNR_j-\EpNR_i\EpNR_j\right)h^{ij}(\tGW)\\
  \hcNR(\tGW) & = \frac{1}{2}\left(\EtNR_i\EpNR_j+\EpNR_i\EtNR_j\right)h^{ij}(\tGW),
\end{align}
\end{subequations}
and
\begin{align}\label{eq:H}
  \sum_{\ell=2}^\infty\sum_{m=-\ell}^\ell H_{\ell m}&(\tGW)\;^{-2}Y_{\ell m}(\tNR,\pNR)
  = \frac{r}{M}\Big(\hpNR(\tGW)-i\hcNR(\tGW)\Big). 
\end{align}
As explained in Sec.~\ref{sec:format}, the data-files are assumed to
contain a compressed time-series of amplitudes and phases per
Eq.~(\ref{ }).  Given an emission direction $(\tNR,\pNR)$,
Eq.~(\ref{eq:H}) yields the GW modes $\hpNR(\tGW)$ and $\hcNR(\tGW)$,
according to the convention Eqs.~(\ref{eq:hNR}).

The gravitational wave data are given in a time-coordinate $\tGW$ of
observers at large distance r.

Let us now turn to a description of the dynamics of the two bodies.
Numerical relativity defines a variety of vectors, the combination of
which defines the instantaneous state of the two bodies.  These are:
The dimensionless spin vectors of the two bodies,
\begin{equation}\label{eq:chi}
  \vec\chi_1(t),\, \vec\chi_2(t);
\end{equation}
the direction from body 2 to body 1,
\begin{equation}
\label{eq:nhat}
  \nNR(t) := \frac{\vec{c}_1 - \vec{c}_2}{|| \vec{c}_1 - \vec{c}_2 ||},
\end{equation}
where $\vec{c}_i$ is the coordinate centre of the horizon of the i-th body;
the direction of the Newtonian orbital angular momentum,
\begin{equation}
  \hat L_N(t).
\end{equation}
We do {\bf not} make any assumption about the relation of the
NR orbital angular momentum vector $\lNR$ relative to the NR
coordinates\footnote{The customary choice of many NR groups is to
  \emph{start} NR simulations with $\lNR=\EzNR$.  Because of
  junk-radiation and precession effects, $\lNR(t)$ will deviate from
  $\EzNR$.  This deviation generally will be very small for
  aligned-spin BBH systems, but may become significant for precessing
  systems.}.
One can further define an orbital frequency
\begin{equation}\label{eq:Omega}
  \vec\Omega_\mathrm{orb}(t) = \nNR(t) \times \frac{d\nNR(t)}{dt}.
\end{equation}

Equations~(\ref{eq:chi})--(\ref{eq:Omega}) are defined in the
strong-field region near the black holes, and are given as a function
of the time-coordinate $t$ employed by the NR simulation in the strong
field regime. 
The time-coordinates $\tGW$ and $t$ are defined at different regions
of the space-time (far-zone vs. near-zone), and their preferred relative
alignment is discussed on page~\ref{tGW-vs-tNR}. Any relation between $\tGW$ and
$t$, however, is ambiguous.  Often NR simulations employ an approximation of
``retarded time'', i.e.
\begin{equation}
  \label{eq:tretarted}
  \tGW = t\big|_{\rm retarded}.
\end{equation}
However, in a dynamical space-time with black holes, two difficulties
arise: First, the horizons are causally disconnected from future null
infinity, so there are no outgoing null-rays that connect the horizons
to the wave-zone.  Secondly, when integrating null-rays ``slightly
outside'' the horizons, the time-delay to infinity will depend on
precisely where the integration was started, as well as on the initial
direction of the null-ray.  Therefore, any relation between $\tGW$ and
$t$ should be viewed as approximate. One should further assume that
different NR groups may use different definitions of $\tGW$ in the
data they compute\footnote{{\tt SpEC}, for example, reports GW-waveforms
extracted at \emph{finite-radius} in terms of the NR coordinate,
$\tGW=t$, whereas extrapolated waveforms are reported using a retarded
time-coordinate with a correction of the rate of flow of time, see
Eqs. (7), (14a) and (14b) of Ref.~\cite{Boyle:2009vi}.}.

\subsubsection{LAL Source-Frame \boldmath$(\ExS, \EyS, \EzS)$}

The LAL source-frame $(\ExS, \EyS, \EzS)$ is defined as follows:
\begin{enumerate}
\item The $\EzS$-axis points along the orbital angular momentum of the binary,
  \begin{equation}\label{eq:EzS}
    \EzS\equalref \lNR.
  \end{equation}
\item The $\ExS$-axis points along the vector $\nNR$ pointing from the second
  to the first body,
  \begin{equation}\label{eq:ExS}
    \ExS\equalref\nNR.
  \end{equation}
\item The third vector $\EyS=\EzS\times\ExS$ completes the triad.
\end{enumerate}
The symbol $\equalref$ indicates that the
respective equation is only required at the reference epoch.  The
reference epoch can be unambiguously specified by a reference-time
$t_{\rm ref}$.  One can also specify a reference \emph{orbital}
frequency $\Omega_{\rm orb,ref}$, and infer $t_{\rm ref}$ via
Eq.~(\ref{eq:Omega}), $\Omega_{\rm orb}(t_{\rm ref})=\Omega_{\rm orb, ref}$.  If the
reference epoch is desired to be specified in terms of a
\emph{gravitational wave}--frequency, then the gravitational wave time
$\tGW$ needs to be related to the time-coordinate of the black hole
dynamics, $t$.  As discussed in the context of
Eq.~(\ref{eq:tretarted}), such an identification is ambiguous and
holds only approximately.

Equations~(\ref{eq:EzS}) and~(\ref{eq:ExS}) define the source-frame at
the reference epoch only.  They are chosen such that the
spin-components ({\tt Sx1, Sy1, Sz1}) and ({\tt S2x, S2y, S2z}) have
coordinate-invariant meaning: {\tt Sx1} is the projection of
$\vec\chi_1$ onto $\nNR$, {\tt Sz1} is the projection of $\vec\chi_1$
onto $\lNR$, etc.

The source-frame does not rotate as the binary evolves.  Specifically,
the rotation between source-frame and wave-frame described below is
constant in time.

A different reference epoch would lead to a different source-frame,
related by some rotation.  If the reference epoch is shifted by a
small amount (comparable to the orbital time-scale), $\ExS$ and $\EyS$
would rotate with the binary.  On the precession time-scale $\EzS$
would change.

The source frame has no deep intrinsic, geometric significance.  It is
merely a vehicle to describe the spin-projections onto intrinsic
geometric directions, and the basis-vectors $(\ExS, \EyS, \EzS)$ will
be convenient when writing down the transformation to the wave-frame.

\subsubsection{Intrisinc parameters of a binary}

Given a reference epoch, a binary is specified by the following ten
numbers:
\begin{itemize}
\item Two masses $m_1$, $m_2$.
\item Two spin-vectors $\vec\chi_1$, $\vec\chi_2$, specified through 
   the projections of the spin-vectors onto $\lNR$, $\nNR$
  and the third basis-vector (at reference time).
\item Eccentricity $e$.
\item Mean anomaly $\meanAnomaly$.  The mean anomaly is defined in terms of the current time $t$, the time of last periapsis passage $T_{\rm previous}$, and the time of next periapsis passage $T_{\rm next}$: 
\begin{equation}
\label{eq:anomaly}
\meanAnomaly=2\pi\frac{t-T_{\rm previous}}{T_{\rm next}-T_{\rm previous}}.
\end{equation}
\end{itemize}
Eccentricity should be defined somehow ``near'' the reference epoch.
A precise definition (if possible at all) is left to the future.  For
low eccentricity orbits, there is a competition between radiation
reaction driven inspiral (which results in a slightly negative average
radial velocity), and the oscillatory radial motion due to
eccentricity.  For sufficiently small eccentricity, minima in
separation may no longer exist.  If mean anomaly is still needed
despite the quite low eccentricity in such cases, one will have to
define periapsis as minimum of separation compared to a fiducial
smooth inspiral trajectory.

\subsubsection{Wave-Frame \boldmath$(\ExW, \EyW, \EzW)$}
\label{sec:WaveFrame}

The wave-frame is adopted to the direction of the observer
(i.e. Earth), such that its $\EzW$-axis points toward the observer.
$\ExW$ and $\EyW$ represent basis-vectors orthogonal to the
line-of-sight, i.e. they span the plane of the sky.
The wave-frame is completely specified by three angles:
\begin{enumerate}
  \item The angle $\phiRef$ between line of ascending node and $\nNR$
    (at the reference time). 
  \item The inclination $\iota$, i.e. the angle between orbital
    angular momentum $\lNR=\EzS$ and the line-of-sight $\EzW$.
  \item The angle $\Omega$ between the $\ExW$-axis and the line of the
    ascending node (at the reference time).
\end{enumerate}
These three angles happen to be the Euler angles of the rotation from
the wave-frame to the source-frame. In the LAL wave-frame, the gravitational wave modes are defined as
\begin{subequations}
  \label{eq:hW}
  \begin{align}
       \hpW &= \frac{1}{2}\left(\ExW_i\ExW_j-\EyW_i\EyW_j\right)h^{ij},\\
       \hcW &= \frac{1}{2}\left(\ExW_i\EyW_j+\EyW_i\ExW_j\right)h^{ij},
  \end{align}
\end{subequations}
where the superscript 'W' indicates the LAL wave-frame.  The angle
$\Omega$ rotates $\ExW$ and $\EyW$ into each other. By
definition of Eq.~(\ref{eq:hW}), this merely rotates the polarization of
the GW modes, and so $\Omega$ is fully degenerate with the GW
polarization.

Note that the phase-angle $\phiRef$ is specified without regard to the
location of periapsis of the binary (this differs from the previous LAL
convention).  This new definition decouples the specification of
periapsis location and the specification of orbital phase, and avoids
the ambiguity that would arise in the zero-eccentricity limit of
definitions involving periapsis. Indeed, for fixed $\phiRef$ and fixed
$\meanAnomaly$, as eccentricity approaches zero, the waveforms will
approach the identical circular form, independent of the value of
$\meanAnomaly$.

\subsection{Relationship between frames}
\label{sec:Relations}

\subsubsection{From source-frame to wave-frame}

The relation between source-frame and wave-frame can be easily derived
following the three rotations that rotate one frame into the other:
Beginning in the source-frame, $\ExS, \EyS, \EzS$, we first apply the rotation
by $\phiRef$ around $\EzS$, which yields a frame with basis-vectors $\hat p, \hat q, \EzS$:
\begin{subequations}
  \label{eq:phiRef-Rotation1}
\begin{align}
  \hat p &= \cos\phiRef\,\ExS - \sin\phiRef\,\EyS,\\
  \hat q &= \sin\phiRef\,\ExS+\cos\phiRef\,\EyS,\\
  \EzS & = \EzS.
\end{align}
\end{subequations}
The inverse rotation is
\begin{subequations}
  \label{eq:phiRef-Rotation2}
\begin{align}
  \ExS &= \;\;\;\cos\phiRef\,\hat p + \sin\phiRef\,\hat q,\\
  \EyS &= -\sin\phiRef\,\hat p+\cos\phiRef\,\hat q,\\
  \EzS & = \EzS.
\end{align}
\end{subequations}
The vectors $\hat p, \hat q$ form an orthonormal basis of the
$\ExS$-$\EyS$ plane, such that $\hat p$ points in the direction of
ascending node. \\
Next, we rotate around the line of the ascending node by the inclination $\iota$,
resulting in basis-vectors
\begin{subequations}
  \label{eq:iota-Rotation1}
\begin{align}
  \hat P&=\hat p,\\
  \hat Q&=\cos\iota\,\hat q - \sin\iota\,\EzS,\\
  \hat Z&=\sin\iota\,\hat q+\cos\iota\,\EzS.
\end{align}
\end{subequations}
 $\hat P$ and $\hat Q$ form an orthonormal basis of the $\ExW$-$\EyW$-plane 
 with $\hat P$ pointing in the direction of the ascending node.
The inverse rotation is
\begin{subequations}
  \label{eq:iota-Rotation2}
\begin{align}
  \hat p&=\hat P,\\
  \hat q&=\;\;\;\cos\iota\,\hat Q + \sin\iota\,\EzW,\\
  \EzS &=-\sin\iota\,\hat Q+\cos\iota\,\EzW.
\end{align}
\end{subequations}
The final rotation rotates $\hat P$ and $\hat Q$
around $\EzW$ into $\ExW,\EyW$:
\begin{subequations}
  \label{eq:Omega-Rotation1}
\begin{align}
  \ExW&=\cos\Omega\,\hat P-\sin\Omega\,\hat Q,\\
  \EyW&=\sin\Omega\,\hat P+\cos\Omega\,\hat Q,\\
  \EzW&=\EzW.
\end{align}
\end{subequations}
The inverse rotation is
\begin{subequations}
  \label{eq:Omega-Rotation2}
\begin{align}
  \hat P&=\;\;\;\cos\Omega\,\ExW+\sin\Omega\,\EyW,\\
  \hat Q&=-\sin\Omega\,\ExW+\cos\Omega\,\EyW,\\
  \EzW&=\EzW.
\end{align}
\end{subequations}
Substituting Eqs.~(\ref{eq:phiRef-Rotation1}),~(\ref{eq:iota-Rotation1}) and~(\ref{eq:Omega-Rotation1}) into each other, one obtains the entire transformation:
\begin{subequations}
  \label{eq:Source-To-Wave}
  \begin{align}
    \ExW=& \left(\cos\Omega\cos\phiRef-\sin\Omega\cos\iota\sin\phiRef\right)\ExS
    \nonumber \\
    & + \left(-\cos\Omega\sin\phiRef-\sin\Omega\cos\iota\cos\phiRef\right)\EyS
    \nonumber\\
    & + \sin\Omega\sin\iota\,\EzS,\\
\EyW=& \left(\sin\Omega\cos\phiRef+\cos\Omega\cos\iota\sin\phiRef\right)\ExS
    \nonumber \\
    & + \left(-\sin\Omega\sin\phiRef+\cos\Omega\cos\iota\cos\phiRef\right)\EyS
    \nonumber\\
    & - \cos\Omega\sin\iota\,\EzS,\\
\label{eq:Z_from_z}
\EzW=&\sin\iota\sin\phiRef\,\ExS+\sin\iota\cos\phiRef\,\EyS + \cos\iota\,\EzS.
  \end{align}
  \end{subequations}
The inverse transformation is obtained from
Eqs.~(\ref{eq:phiRef-Rotation2}), (\ref{eq:iota-Rotation2})
and~(\ref{eq:Omega-Rotation2}):
\begin{subequations}
  \label{eq:Wave-To-Source}
  \begin{align}
    \ExS=& \left(\cos\Omega\cos\phiRef-\sin\Omega\cos\iota\sin\phiRef\right)\ExW
    \nonumber \\
    & + \left(\sin\Omega\cos\phiRef+\cos\Omega\cos\iota\sin\phiRef\right)\EyW
    \nonumber\\
    & + \sin\iota\sin\phiRef\,\EzW,\\
\EyS=& \left(-\cos\Omega\sin\phiRef-\sin\Omega\cos\iota\cos\phiRef\right)\ExW
    \nonumber \\
    & + \left(-\sin\Omega\sin\phiRef+\cos\Omega\cos\iota\cos\phiRef\right)\EyW
    \nonumber\\
    & + \sin\iota\cos\phiRef\,\EzW,\\
\label{eq:z_from_Z}
    \EzS=&\sin\Omega\sin\iota\,\ExW-\cos\Omega\sin\iota\,\EyW + \cos\iota\,\EzW.
  \end{align}
  \end{subequations}
Equation~(\ref{eq:z_from_Z}) shows that $\Omega$ determines the
direction of $\lNR=\hat z$ on the plane of the sky.  For $\Omega=0$,
$\lNR$ lies in the $\EyW$-$\EzW$ plane, and for $\Omega=\pi/2$, it lies
in the $\ExW$-$\EzW$ plane. This latter choice ($\Omega=\pi/2$) is already 
respected by many LAL
waveform models.  Therefore, in the absence of a reason to do
otherwise, all waveform models should default to $\Omega=\pi/2$.  \\
With this recommended default $\Omega=\pi/2$,
Eqs.~(\ref{eq:Source-To-Wave}) simplify to
\begin{subequations}
  \label{eq:Source-To-Wave-Omega0}
  \begin{align}
    \ExW=& -\cos\iota\sin\phiRef\,\ExS-\cos\iota\cos\phiRef\,\EyS\,+\sin\iota\,\EzS,\\
    \EyW=&\quad\quad\;\;\; \cos\phiRef\,\ExS\qquad\,-\sin\phiRef\,\EyS,\\
\label{eq:Z_from_z-Omega0}
\EzW=&\;\;\;\;\,\sin\iota\sin\phiRef\,\ExS+\sin\iota\cos\phiRef\,\EyS\, + \cos\iota\,\EzS.
  \end{align}
  \end{subequations}
The inverse transformation (\ref{eq:Wave-To-Source}) simplifies to 
\begin{subequations}
  \label{eq:Wave-To-Source-Omega0}
  \begin{align}
    \ExS=& -\cos\iota\sin\phiRef\,\ExW+\cos\phiRef\,\EyW
    + \sin\iota\sin\phiRef\,\EzW,\\
    \EyS=& -\cos\iota\cos\phiRef\,\ExW-\sin\phiRef\,\EyW
    + \sin\iota\cos\phiRef\,\EzW,\\
    \EzS=&\qquad\quad\;\,\sin\iota\,\ExW \qquad\qquad\qquad\;+\cos\iota\,\EzW.
  \end{align}
  \end{subequations}

\subsubsection{From NR-frame to wave-frame}

Equations~(\ref{eq:Source-To-Wave}) and
(\ref{eq:Source-To-Wave-Omega0}) are of particular importance.  By
definition, the source-frame basis-vectors are trivially related to
vectorial quantities of the compact binary dynamics:
\begin{subequations}
  \begin{align}
    \ExS&\equalref \nNR,\\
    \EyS&\equalref (\lNR\times\nNR),\\
    \EzS&\equalref \lNR.
  \end{align}
\end{subequations}
Therefore, if the dynamics vectors $\lNR$ and $\nNR$ are known in any
coordinate system, e.g. the NR coordinates, then
Eqs.~(\ref{eq:Source-To-Wave}) yield the wave-frame basis-vectors in those
coordinates.  Specifically, for $\Omega=\pi/2$,
Eqs.~(\ref{eq:Source-To-Wave-Omega0}) yield
\begin{subequations}
  \begin{align}
    \ExW &\equalref-\cos\iota\left[\sin\phiRef\,\nNR +\cos\phiRef\,\lNR\times\nNR\right]+\sin\iota\lNR,\\
    \EyW &\equalref \qquad\quad\;\,\cos\phiRef\,\nNR -\sin\phiRef\,\lNR\times\nNR,\\
\label{eq:Z}
    \EzW&\equalref\;\;\;\; \sin \iota\left[\sin\phiRef\,\nNR +\cos\phiRef\,\lNR\times\nNR\right]
    +\cos \iota \,\lNR.
  \end{align}
\end{subequations}

Let us consider next the transformation of the GW strain polarizations
from the generic (NR) frame Eq.~(\ref{eq:NRspherical}) to the
wave-frame.  $\ExW$ and $\EyW$ are orthogonal to the direction of
propagation $\EzW=\ErNR$ of the gravitational wave.  Therefore,
$(\EtNR, \EpNR)$ can be rotated into $(\ExW, \EyW)$ through a rotation
by an angle $\alpha$:
\begin{subequations}\label{eq:alpha}
\begin{align}
\ExW & = \cos\alpha\,\EtNR - \sin\alpha\,\EpNR,\\
\label{eq:alphaY}
\EyW & = \sin\alpha\,\EtNR +\cos\alpha\,\EpNR.
\end{align}
\end{subequations}
Substituting Eqs.~(\ref{eq:alpha}) into Eqs.~(\ref{eq:hW}) we can compute $\hpW$ and $\hcW$ in terms of $(\EtNR,\EpNR)$.  Comparing further with Eqs.~(\ref{eq:hNR}), we find:
\begin{subequations}
\label{eq:PolarizationTrafo}
\begin{align}
  \hpW(\tGW) &= \cos\left(2\alpha\right)\,\hpNR(\tGW) -\sin\left(2\alpha\right)\,\hcNR(\tGW),\\
  \hcW(\tGW) &= \sin(2\alpha)\,\hpNR(\tGW)+\cos\left(2\alpha\right)\,\hcNR(\tGW).
\end{align}
\end{subequations}
Not surprising, the GW polarizations in the wave-frame are obtained
from those in the NR-frame by a rotation of $2\alpha$.

The wave-frame $\ExW, \EyW$ depend on $\Omega$, and therefore $\hpW$
and $\hcW$ also depend on $\Omega$.  We can make this dependence
explicit by resorting to the intermediate vectors $\hat P$ and
$\hat Q$.  These are also orthogonal to $\EzW$, therefore they, too,
can be obtained from $\EtNR$ and $\EpNR$ by a rotation:
\begin{subequations}
\begin{align}
\hat P & = \cos\alpha'\,\EtNR - \sin\alpha'\,\EpNR,\\
\hat Q & = \sin\alpha'\,\EtNR + \cos\alpha'\,\EpNR.
\end{align}
\end{subequations}
However, $\hat P$ and $\hat Q$ are independent of $\Omega$ and
therefore, $\alpha'$ is independent of $\Omega$.  Because
$(\ExW, \EyW)$ are rotated by $\Omega$ relative to $(\hat P, \hat Q)$,
we have
\begin{equation}
\alpha = \Omega+\alpha'.
\end{equation}
Because rotations add, we can therefore write
\begin{equation}\label{eq:polarization-rotation}
\left(\begin{aligned}\hpW\\\hcW \end{aligned}\right)
= {\mathbf R}_{2\Omega}\; {\mathbf R}_{2\alpha'} 
\left(\begin{aligned}\hpNR\\\hcNR \end{aligned}\right),
\end{equation}
where ${\mathbf R}_\beta$ denotes a 2x2 rotation matrix,
\begin{equation}
{\mathbf R}_\beta = \left(\begin{aligned}&\cos\beta & -\sin\beta \\ & \sin\beta & \cos\beta\end{aligned}\right).
\end{equation}
Equation~(\ref{eq:polarization-rotation}) thus implies that the
waveform-modes are obtained from the NR-polarizations by (i) applying
an $\Omega$-independent rotation by $2\alpha'$; followed by (ii) a
rotation by $2\Omega$.
For $\Omega=\pi/2$, we have ${\mathbf R}_\pi = -{\mathbf 1}$.
Therefore, between $\Omega=0$ (inclination rotated about X-axis) and
$\Omega=\pi/2$ (inclination rotated about Y-axis), the waveform
polarization pick up precisely an overall minus-sign.

\subsection{Computing GW polarizations in the LAL wave-frame}
\label{sec:NR-LAL-Trafo}
Let us finally write down explicit instructions of how to obtain GW
polarizations in the LAL convention, given NR waveform data.  
Given parameters {\tt i, phiRef} passed into {\tt
XLALSimInspiralChooseTDWaveform}, proceed as follows:

\begin{enumerate}
\item Define $\phiRef=\mbox{phiRef}$, $\iota={\tt i}$.
\item Compute $\EzW_{\rm ref}$ at the reference time by evaluating Eq.~(\ref{eq:Z}).
\item Because $\EzW$ points in the direction of emission of the
gravitational wave, we must have
\begin{equation}
  \EzW_{\rm ref} \equalref
  \left(\begin{gathered}
    \cos\pNR\sin\tNR\\
    \sin\pNR\sin\tNR\\
    \cos\tNR\end{gathered}
    \right).
\end{equation}
From this equality, read off $(\tNR, \pNR)$.  Then compute the
NR-basis vectors $\EtNR, \EpNR$ from Eqs.~(\ref{eq:NRspherical}).
\item If $\Omega=\pi/2$, compute $\sin\alpha$ and $\cos\alpha$ by taking inner products
  of Eq.~(\ref{eq:alphaY}) with $\EtNR$ and $\EpNR$:
  \begin{subequations}
  \begin{align}
    \sin\alpha& = \cos\phiRef\,\nNR\cdot\EtNR- \sin\phiRef\,(\lNR\times\nNR)\cdot\EtNR,\\
    \cos\alpha& = \cos\phiRef\,\nNR\cdot\EpNR - \sin\phiRef\,(\lNR\times\nNR)\cdot\EpNR.
  \end{align}
\end{subequations}
  If $\Omega\neq \pi/2$, compute instead inner products based on
  Eqs.~(\ref{eq:Source-To-Wave}).
\item Substitute $(\tNR, \pNR)$ into Eqs.~(\ref{eq:H})
  and~(\ref{eq:hNR}) to compute $\hpNR(\tGW)$ and $\hcNR(\tGW)$.
\item Compute $\cos 2\alpha=\cos^2\alpha\!-\!\sin^2\alpha$ and $\sin 2\alpha=2\cos\alpha\sin\alpha$.  Substitute into Eqs.~(\ref{eq:PolarizationTrafo}) to compute $\hpW(\tGW)$ and $\hcW(\tGW)$.
\end{enumerate}

\subsubsection{Evaluate spin-consistency in LAL source-frame}

The parameters {\tt S1x, S1y, S1z, S2x, S2y,S2z} passed into {\tt
  LALSimInspiralChooseTDWaveform} are supposed to be the LAL
source-frame parameters, i.e. these parameters should simply be the
projections of $\vec\chi_{1,2}$ onto the source-frame basis-vectors
$(\ExS,\EyS,\EzS)$.  Substituting Eqs.~(\ref{eq:EzS}) and (\ref{eq:ExS}),
one arrives at the following consistency conditions:
\begin{subequations}
\label{eq:SpinConsistency}
  \begin{align}
    \mbox{\tt S1x} &  \equalref \vec\chi_1 \cdot \nNR,\\
    \mbox{\tt S1y} &  \equalref \vec\chi_1\cdot (\lNR\times\nNR),\\
    \mbox{\tt S1z} &  \equalref \vec\chi_1 \cdot \lNR.
  \end{align}
\end{subequations}
The conditions for body 2 are obtained by $1\leftrightarrow 2$.\\


\section{Discussion}
\label{sec:discussion}

With this new infrastructure it is very easy and much less memory intensive to use NR waveforms
directly for data analysis applications. The ``NR\_hdf5'' approximant works much the same as any other approximant
in \texttt{lalsimulation} but there are a few important differences.

First, the user must supply the location of the HDF5 file, a functionality which was already implemented for NINJA, but was not
previously used in \texttt{lalsimulation}. Secondly, the user must be careful to supply the mass ratio and spin values that
are consistent with the NR files, and the spin values have to be specified in the LAL source frame (see Eqs. (\ref{eq:SpinConsistency})). 

The current implementation still suffers from a few caveats and drawbacks. As opposed to the continuous waveform approximants, at the moment the metadata are only referring to the beginning of the waveform and not some reference time, which can be chosen freely. While this is not a problem for aligned-spin binaries, this is a big concern for precessing simulations since various quantities, in particular the spins and the orbital angular momentum, are time-dependent. To fully integrate this desired freedom, additional information needs to be incorporated into the HDF5 files \emph{and} the waveform evaluation functions accordingly. Specifically, one needs the time-series of the vectors determining the geometry of the binary: $\lNR(t), \nNR(t), \chi_(t), \chi_2(t)$. Given these time-series, one can interpolate these four vectors to any reference epoch, and then apply the frame transformations at this reference epoch. These are provided in the formats 2 and 3 and we leave it to future upgrades to \texttt{lalsimulation} to allow for this additional functionality to be fully integrated in the waveform evaluation functions.

\section*{Acknowledgements}
We are grateful to Mark Hannam for many useful discussions and comments throughout the code review.
We also thank Kent Blackburn and James Healy for providing useful comments on the manuscript, and Ian Hinder, Geoffrey Lovelace and Deirdre Shoemaker for input into the metadata discussion.
Many thanks for discussions regarding the frame coordinate transformations to Stas Babak, Jolien Creighton, Michael P\"urrer and Riccardo Sturani.


\bibliography{nrinj}

\end{document}